\documentstyle[prl,aps,preprint,floats]{revtex}

\def\half{\frac{1}{2}}
\input BoxedEPS
\SetRokickiEPSFSpecial  
\HideDisplacementBoxes
\tighten
\begin{document}

\title{Unambiguous One-Loop Quantum Energies of 1+1 Dimensional
Bosonic Field Configurations}

\author{N.~Graham\footnote{graham@mitlns.mit.edu} 
and R.~L.~Jaffe\footnote{jaffe@mitlns.mit.edu}}

\address{{~}\\Center for Theoretical Physics, Laboratory for
  Nuclear Science
  and Department of Physics \\
  Massachusetts Institute of Technology,
  Cambridge, Massachusetts 02139 \\
  {\rm MIT-CTP\#2744\qquad hep-th/9805150}}

\maketitle

\begin{abstract}

We calculate one-loop quantum energies in a renormalizable self-interacting
theory in one spatial dimension by summing the zero-point energies of small
oscillations around a classical field configuration, which need not be a
solution of the classical field equations.  We unambiguously implement
standard perturbative renormalization using phase shifts and the Born
approximation.  We illustrate our method by calculating the quantum energy
of a soliton/antisoliton pair as a function of their separation.  This
energy includes an imaginary part that gives a quantum decay rate and is
associated with a level crossing in the solutions to the classical field
equation in the presence of the source that maintains the
soliton/antisoliton pair.

\end{abstract}

 \narrowtext

\section{Introduction}

The study of exactly soluble 1+1 dimensional problems has yielded many
insights into fundamental problems in field theory. Other 1+1 dimensional
problems cannot be solved exactly, making it important to understand which
properties of exact results will generalize to more generic cases and which
are special to the exactly soluble cases. A particularly rich area is the
study of solitons --- time independent non-linear solutions --- that arise
is interesting theories in both one and three dimensions.  Renormalized
quantum fluctuations can modify a soliton's properties, perhaps even
destabilizing a classically stable solution or stabilizing a classically
unstable one.  The renormalization process is simpler in one dimension than
it is in three, but it still must be approached carefully.  Merely
``canceling the infinities'' is clearly not sufficient when we want to
regard a finite result as a physical prediction of a particular theory
defined under fixed renormalization conditions.

In this Letter we specialize the discussion of \cite{us} to include the
case of 1-dimensional self-interacting theories.  We develop an unambiguous
procedure for applying standard perturbative renormalization methods to the
calculation of the one-loop ``effective energy'' of a time-independent,
spatially symmetric field configuration.  Our method is also practical and
efficient for numerical calculation, which we illustrate by considering a
time-independent family of configurations $\phi(x,x_0)$ that continuously
interpolates between the trivial vacuum $(x_0 = 0)$ and a
soliton/antisoliton pair $(x_0\to\infty)$ in $\phi^4$ theory.  The energy
of these configurations can include an imaginary part, which we show
measures the instability of the system to quantum fluctuations.  We also 
consider the exactly soluble cases such as the $x_0\to\infty$ limit of the
soliton/antisoliton pair, and obtain standard results that are free of
cutoff ambiguities.  Throughout, we will be able to gain insight into
these field-theoretic problems by understanding them in terms of
physical quantities like phase shifts and bound state energies.

The present work can be regarded in several ways.  On the one hand,
one-dimensional models provide very clean pedagogical examples of the
general methods developed in \cite{us}.  On the other hand, quantum
energies of one dimensional solitons are interesting in their own right, as
\cite{vN}, \cite{Naculich1d}, and \cite{Bordag} attest.  Our methods allow
us to resolve questions that are obscure if one can only analyze quantum
fluctuations about {\it solutions} to the classical equations of motion. 
Our results on the quantum fluctuations about the $\phi^4$ kink are not new
--- they coincide with the well known results of Dashen, Hasslacher and
Neveu \cite{DHN}.  However, our method displays interesting features like
the correlation between imaginary frequency small oscillations and
level crossing in the space of solutions to the equations of motion,
and provides a simple method for avoiding ambiguities.

\section{Formalism}

To illustrate our method, we will consider a standard $\phi^4$ theory
with spontaneous symmetry breaking and a source $J(x)$.  The action is
\begin{equation}
S[\phi] = \frac{m^2}{\lambda} \int 
\left(\half(\partial_\mu \phi)^2 -  \frac{m^2}{8}(\phi^2 - 1)^2
  - J(x)\phi + {\cal L}_{\rm ct} \right) \, d^2x
\end{equation}
where ${\cal L}_{\rm ct}$ is the counterterm Lagrangian.  We have
rescaled the field $\phi$ somewhat unconventionally in order to make
explicit the correspondence between the powers of $\lambda$ and the powers
of $\hbar$.  Classical terms will go as $\frac{1}{\lambda}$, the one-loop
terms we compute will go as $\lambda^{0}$, and higher loops will
contribute with higher powers of $\lambda$.  The mass of fluctuations
around the trivial vacua $\phi(x)=\pm 1$ is $m$, and we define the
potential $U(\phi)=\frac{m^2}{8}(\phi^2 - 1)^2$.

We consider a fixed field configuration $\phi_0(x)$.  Initially we will
assume that $\phi_0(x) = \phi_0(-x)$, which restricts us to the
topologically trivial sector of the theory.  However, we will see that our
method works equally well for the case of $\phi_0(x) = -\phi_0(-x)$, so
that in fact we can deal with configurations with any topology as long as
$U(\phi_0)$ has reflection symmetry.  We adjust the source so that $\phi_0$ is
a stationary point of the action, which means that $J$ then solves the
equation
\begin{equation}
J(x) = \frac{d^2 \phi_0}{dx^2} - \half m^2 (\phi_0^3 - \phi_0).
\label{Jeq}
\end{equation}
We can always solve this equation for $J$, but there is no guarantee that the
$J$ we find will always correspond to a unique $\phi_0$, as we will see later.
If $\phi_0$ is a solution to the equations of motion, of course the source
will be zero.

We would like to consider the leading quantum correction to the classical
energy of this configuration, which we can represent as the sum of
the zero-point energies of the normal modes of small oscillations
around $\phi_0$.  Writing $\phi = \phi_0 + \eta$, the normal modes are
solutions of
\begin{equation}
-\frac{d^2 \eta}{dx^2} + (V(x)+m^2)\eta = \omega^2 \eta
\label{etaeq}
\end{equation}
with $V(x) = U''(\phi_0(x)) - m^2 = \frac{3}{2} m^2 (\phi_0^2(x) - 1)$.
The quantum change in energy in going from the trivial vacuum to
$\phi_0(x)$ is then
\begin{equation}
{\cal E}[\phi_0] = \half(\sum \omega - \sum \omega_0) + {\cal E}_{\rm ct}
= \Delta E + {\cal E}_{\rm ct}
\end{equation}
where $\omega_0$ are the free solutions (with $\phi_0^2(x) = 1$), and
${\cal E}_{\rm ct}$ is the contribution from the counterterms.

Note that it is possible that some of the values of $\omega^2$ will be
negative, so that in these directions our stationary point is a local
maximum rather than a minimum of the action.  These solutions will add an
imaginary part to the energy, which we can interpret via analytic
continuation as giving the decay rates through the unstable modes
\cite{Cole}. If there is a direction in field space in which small
oscillations lower the energy, we should be able to keep going in that
direction and arrive at a lower minimum.  Later, we will see explicitly
that the appearance of unstable modes coincides with the existence of a second
solution to the $\phi_0$ equation with the same $J$ and lower energy.

We would like to rewrite the sum over zero-point energies as an integral
over phase space of the product of the energy and the density of
states.  We can then break this integral into a sum over bound states
and an integral over a continuum, representing the latter in terms of
phase shifts.  In order to do so, however, we must review the
peculiarities of Levinson's theorem in one dimension.  For more
details on these results see \cite{Schiff} and \cite{lev1}.  For a
symmetric $V(x)$, we can divide the continuum states into symmetric
and antisymmetric channels, and then calculate the phase shift as a
function of $k$ separately for each channel, where $\omega^2 = k^2 +
m^2$. The antisymmetric channel is completely equivalent to the $l =
0$ case in three dimensions, so we have
\begin{equation}
\delta_{\rm A}(0) = n_{\rm A} \pi
\end{equation} where $n_{\rm A}$ is the number of antisymmetric bound
states.  However, we must be careful in dealing with the special case of a
state exactly at $k = 0$.  In this case the solution to eq.~(\ref{etaeq})
with $k = 0$ goes asymptotically to a constant as $x\to\infty$, as opposed
to the generic case where the $k=0$ solution goes to a constant plus linear
terms in $x$.  Just as in the $l = 0$ case in three dimensions, this state
contributes $\half$ to $n_{\rm A}$.  We will refer to such
states as ``half-bound states.''

In the symmetric channel, Levinson's theorem becomes
\begin{equation}
\delta_{\rm S}(0) = n_{\rm S} \pi - \frac{\pi}{2}
\end{equation}
where a bound state at $k = 0$ contributes $\half$ to $n_{\rm S}$, just as
in the antisymmetric case.  We can see the importance of getting the
half-bound states right by looking at the free case: then the phase shift
is zero everywhere, and the  right-hand side is zero because the free case
has a half-bound state (the wavefunction $\psi = \rm{constant}$).  The
situation is equally subtle for reflectionless potentials, all of which
have half-bound states and $\delta_{\rm S}(0) + \delta_{\rm A}(0)$ equal to
an integer times $\pi$.

We are now ready to rewrite the change in the zero-point energies in terms
of phase shifts.  Letting $E_j$ be the bound state energies (again with $k = 0$
bound states contributing with a $\half$), $\rho(k)$ be the density of states
and $\rho_0(k)$ be the free density of states, we have
\begin{eqnarray}
\Delta E
&=& \half \sum_j E_j - \frac{m}{4} + \int_{0}^{\infty} \frac{dk}{2\pi}
\omega(k) (\rho(k) - \rho_0(k)) \nonumber \\
&=& \half \sum_j E_j - \frac{m}{4} + \int_{0}^{\infty} \frac{dk}{2\pi}
\omega(k) \frac{d}{dk}(\delta_{\rm A}(k) + \delta_{\rm S}(k))
\label{unrenorm}
\end{eqnarray}
where $\omega(k)=\sqrt{k^2+m^2}$ and we have used
\begin{equation}
\rho(k) = \rho_0(k) + \frac{1}{\pi} 
\frac{d}{dk}(\delta_{\rm A}(k) + \delta_{\rm S}(k)).
\end{equation}
Note that the $\frac{m}{4}$ term subtracts the contribution of the
free half-bound state.

Eq.~(\ref{unrenorm}) is divergent (the phase shifts fall as
$\frac{1}{k}$ for $k \to\infty$), which is what we should expect since it
includes the divergent contribution from the tadpole graph without the
divergent contribution from the counterterms that cancels it.

To avoid infrared problems later, we first use Levinson's theorem to
compute the change in particle number, which is given by
\begin{equation}
0 = \sum_j 1 + \int_{0}^{\infty} \frac{dk}{\pi}
\frac{d}{dk}(\delta_{\rm A}(k) + \delta_{\rm S}(k)) - \half
\end{equation}
where again, half-bound states are counted with a $\half$ in
the sum over $j$ and the $-\half$ comes from the contribution of the
free half-bound state.  Subtracting $\frac{m}{2}$ times this equation from
eq.~(\ref{unrenorm}), we  have
\begin{equation}
\Delta E = \half\sum_j (E_j-m) + \int_{0}^{\infty} \frac{dk}{2\pi}
(\omega(k)-m) \frac{d}{dk}(\delta_{\rm A}(k) + \delta_{\rm S}(k)).
\label{unrenormprime}
\end{equation}

Next, we follow \cite{us} and subtract the first Born approximation to
eq.~(\ref{unrenormprime}), which corresponds exactly to the tadpole graph. 
We must then add it back in using ordinary renormalized perturbation
theory.  However, in 1+1 dimensions we can adopt the simple renormalization
condition that the counterterms cancel the tadpole graph and perform no
additional finite renormalizations beyond this cancellation.  With this
choice, there is then nothing to add back in.

The first Born approximation is given by
\begin{eqnarray}
\delta^{(1)}_{\rm S}(k) &=& -\frac{1}{k}\int_{0}^{\infty} V(x) \cos^2 kx
\, dx
\nonumber \\
\delta^{(1)}_{\rm A}(k) &=& -\frac{1}{k}\int_{0}^{\infty} V(x) \sin^2 kx
\, dx.
\label{Borneqn}
\end{eqnarray}
Notice that the sum of these two depends on $V(x)$ only
through the quantity $\langle V\rangle = \int_{0}^{\infty} dx V(x)$, so we
can indeed cancel the tadpole contribution with available counterterms.

Subtracting the first Born approximation, we have
\begin{equation}
{\cal E}[\phi_0] = \half\sum_j (E_j-m) + \int_{0}^{\infty} \frac{dk}{2\pi}
(\omega(k)-m) \frac{d}{dk}(\delta_{\rm A}(k) + \delta_{\rm S}(k) +
\frac{\langle V\rangle}{k})
\label{renorm1}
\end{equation}
which is completely finite and well-defined, since the integrand goes like
$\frac{1}{k^2}$ for $k\to\infty$ and goes to a constant at $k = 0$.
In particular, we are free to integrate by parts, giving an
expression that will be easier to deal with computationally,
\begin{equation}
{\cal E}[\phi_0] = \half\sum_j (E_j-m) - \int_{0}^{\infty} \frac{dk}{2\pi}
\frac{k}{\omega(k)} (\delta_{\rm A}(k) + \delta_{\rm S}(k) +
\frac{\langle V\rangle}{k}).
\label{renorm2}
\end{equation}

The subtraction has canceled the leading $\frac{1}{k}$ behavior of the
phase shift, leaving an integral that is finite in the ultraviolet.  It
must, since in 1+1 dimensions, eliminating the tadpole is sufficient to
render the theory finite. (In 3+1 dimensions, we can decompose the problem
into a sum of 1-dimensional  problems, labeled by the angular momentum
$\ell$.  We then need two subtractions:  the first subtraction renders each
1-dimensional problem finite, and the second renders the sum over $\ell$
finite.)

Our use of Levinson's theorem to ``subtract'' $\Delta E$ in the infrared
(replacing eq.~(\ref{unrenorm}) by
eq.~(\ref{unrenormprime})) avoided a subtlety of the Born approximation in
one dimension that does not occur in higher dimensions:  as $k\to 0$, the
symmetric contribution from eq.~(\ref{Borneqn}) introduces a spurious
infrared divergence, since it goes like $\frac{1}{k}$.  (Each higher
dimension adds a power of $k$ near $k=0$.  We can see this in the Feynman
diagram calculation, where these powers of $k$ come from the
measure.)

\section{Applications}

We can now use eq.~(\ref{renorm2}) to calculate quantum energies for specific
field configurations.  Knowing that our model has a ``kink''
soliton solution $\phi_0(x) = \tanh\frac{mx}{2}$, we will consider a
family of field configurations that continuously interpolate between the
trivial vacuum and a soliton/antisoliton pair,
\begin{equation}
\phi_0(x,x_0) = \tanh\frac{m}{2}(x+x_0) - \tanh\frac{m}{2}(x-x_0) - 1,
\end{equation}
with $2x_0$ measuring approximately the separation
between the soliton and antisoliton. Unlike the kink, these configurations
are not solutions of the equations of motion, except in the $x_0\to 0$ and
$x_0\to \infty$ limits.  Thus we will need to to introduce a source that
will vanish in these limits, and we must analyze the stability questions we
raised earlier.

For $x_0$ very small, we simply have a small attractive perturbation from
the trivial vacuum held in place by a small source, which we would not
expect to introduce any instabilities.  In terms of the scattering
problem, for small $x_0$, the potential is too weak to bind a state with
a binding energy greater than $m$, which would give an imaginary
eigenvalue.

For $x_0$ very large, we have a widely separated
soliton/antisoliton pair.  We know from translation invariance that a
single soliton has a mode with $\omega^2 = 0$.  Since this zero mode
corresponds to a nodeless wavefunction, it is the lowest energy mode.  The
soliton/antisoliton pair has two translation modes that will mix,
giving a symmetric eigenstate with a slightly lower energy and an
antisymmetric eigenstate with a slightly higher energy.  Thus we expect to
find a single symmetric mode with $\omega^2 < 0$, which will contribute an
imaginary part to ${\cal E}[\phi_0]$.  The imaginary part gives the rate
for our field configuration to decay through this mode toward the trivial
vacuum.  As $x_0\to\infty$, we should find that the imaginary part goes to
zero, and the real part goes to twice the energy of a single soliton, which
we compute exactly below. 

According to this analysis, there should be a finite, nonzero value of
$x_0$ where the imaginary eigenvalue first appears.  At this point,
the field becomes unstable with respect to small perturbations in some
direction in field space.  Therefore, the energy must have a lower
minimum that we can reach by moving in that direction.
This configuration $\psi_0(x,x_0)$ is a second stationary point of the
action with the same source, which crosses $\phi_0$ at the value of
$x_0$ where the imaginary part for the energy appears.
(For smaller values of $x_0$, this solution still exists but has
higher energy.)  In terms of the scattering problem, this crossing
appears when the potential has a bound state with $\omega^2 = 0$.  We
can identify this state explicitly:  since $\phi_0$ and $\psi_0$
satisfy eq.~(\ref{Jeq}) with the same $J$, as we approach the
crossing, the wavefunction $\eta(x) = \psi_0(x,x_0)-\phi_0(x,x_0)$
becomes a solution to eq.~(\ref{etaeq}) with $\omega^2 = 0$.

We have carried out this computation numerically, and find results that
agree with all of these expectations.  To compute the antisymmetric phase
shift, we use the method of \cite{us} to write
\begin{equation}
\delta_{\rm A}(k) = - 2 \: {\rm Re} \: \beta(k,0)
\end{equation}
where $\beta(k,x)$ is 0 at $x=\infty$ and satisfies
\begin{equation}
-i\beta'' + 2 k\beta' + 2 (\beta')^2 + \half V(x) = 0,
\end{equation}
with prime denoting differentiation with respect to $x$.  To obtain the
symmetric phase shift, we can follow the same derivation, but instead of
imposing $\psi(0) = 0$ on the wavefunction, we instead impose $\psi'(0) = 0$,
giving
\begin{equation}
e^{2i\delta_{\rm S}} = \frac{e^{2i\beta}}{e^{-2i\beta^\ast}}
\frac{k+2\beta'}{k+2\beta'^\ast}
\end{equation}
with $\beta$ the same as above, so that
\begin{equation}
\delta_{\rm S}(k) = \delta_{\rm A}(k) -
\arctan(\frac{2 \: {\rm Im} \: \beta'(k,0)}{k + 2\: {\rm Re} \: \beta'(k,0)}).
\end{equation}

\begin{figure}[htbp]
$$
\BoxedEPSF{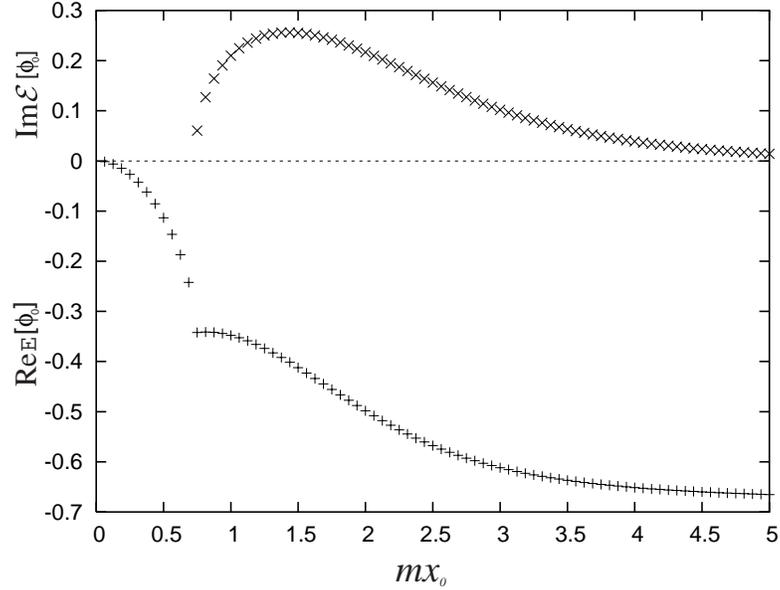 scaled 600}  
$$
\caption{The real and imaginary parts of ${\cal E}[\phi_0]$, with ${\cal
E}$ in units of $m$, for $x_0$ increasing from $0$ to $\frac{5}{m}$.}
\label{figure1}
\end{figure}

\begin{figure}[htbp]
$$
\BoxedEPSF{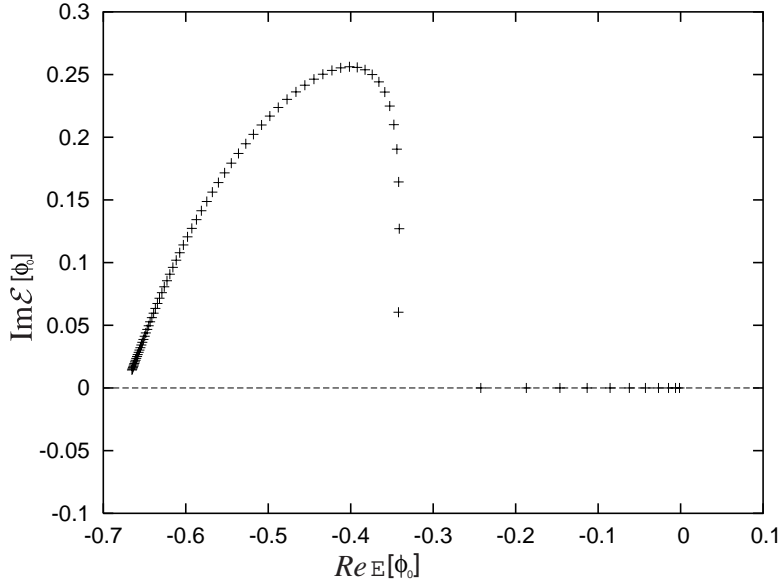 scaled 600}  
$$
\caption{The trajectory of ${\cal E}[\phi_0]$ in the complex plane,
with ${\cal E}$ in units of $m$, for $x_0$ increasing from $0$ to
$\frac{5}{m}$, in steps of 0.0625/m.}
\label{figure1a}
\end{figure}

Figure~\ref{figure1} shows the real and imaginary parts of ${\cal E}[\phi_0]$
as a functions of $x_0$, starting from the
origin at $x_0 = 0$.  When $x_0\approx \frac{0.75}{m}$, a single imaginary
eigenvalue appears and ${\cal E}[\phi_0]$ becomes complex.  For $x_0$
large, ${\cal E}[\phi_0]$ approaches $2(\frac{1}{4
\sqrt{3}} - \frac{3}{2\pi})$, twice the standard result for a single
soliton.  Figure~\ref{figure1a} shows the path of the energy in the complex
plane parametrically as a function of the separation.  The actual trajectory
of ${\cal E}[\phi_0]$ in the complex plane has little significance, since
it depends in detail on the functional form of $\phi(x,x_0)$.  However, the
general features --- beginning at the origin, moving up the real axis, out
into the complex plane, and finally asymptotically to the real two-solution
value --- are characteristics of any $\phi_0$ that begins at the vacuum and
ends at a well separated kink and antikink.

\begin{figure}[htb]
$$
\BoxedEPSF{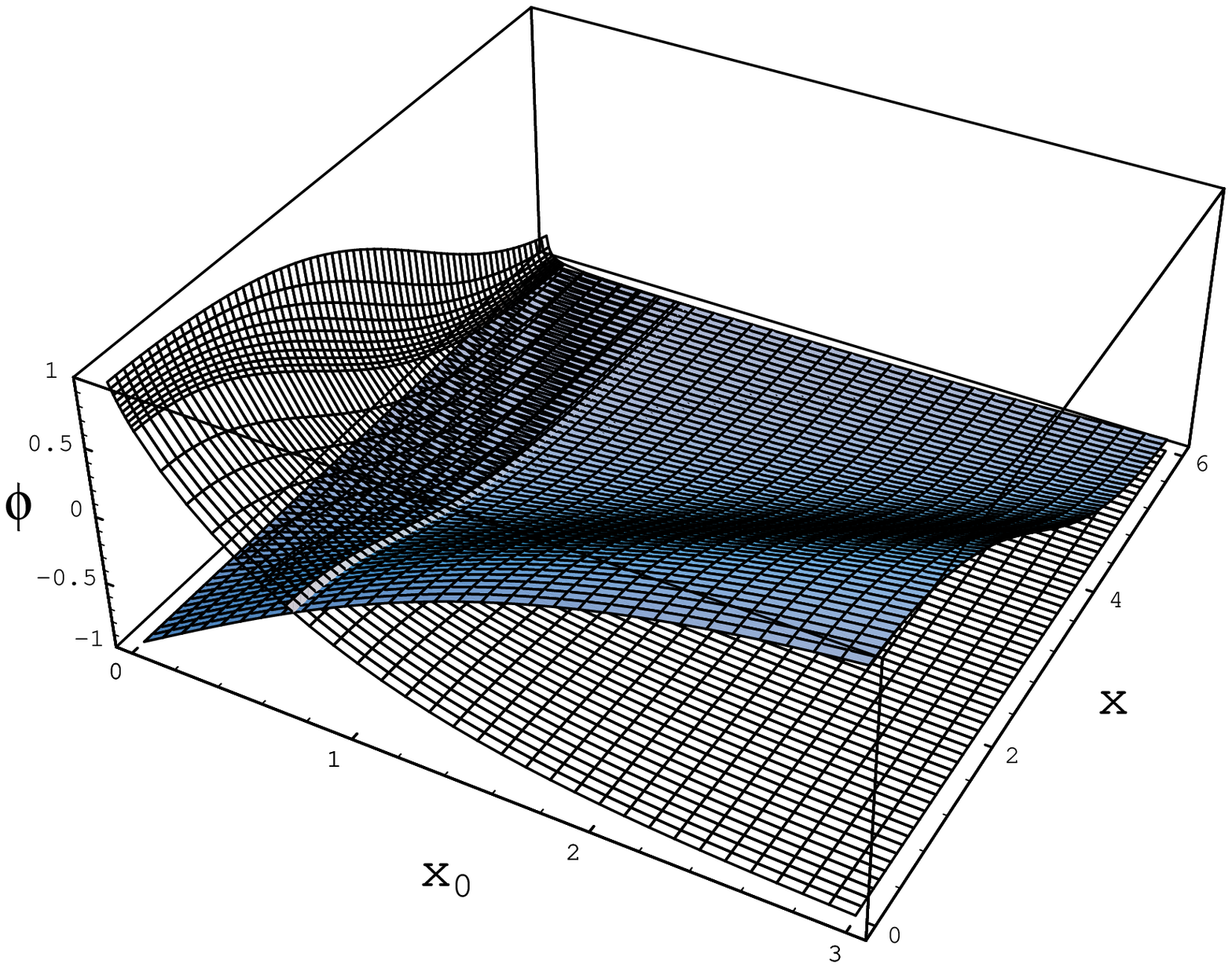 scaled 800}  
$$
\caption{Solutions to the equations of motion with a source given by
eq.~(\ref{Jeq}), as functions of $x$ and $x_0$ in units of $\frac{1}{m}$.  
The shaded graph is $\phi_0(x,x_0)$, which is guaranteed to be a
solution by the construction of $J$.  The unshaded graph gives the
second solution $\psi_0(x,x_0)$.  (A finer mesh is used for this graph
between $x_0 = 0$ and $x_0 = \frac{0.1}{m}$ in order to illustrate its
behavior in this region.) 
}
\label{figure2}
\end{figure}

Fig.~\ref{figure2} shows $\phi_0(x,x_0)$ and the second solution to
the equations of motion with the same $J$, $\psi_0(x,x_0)$.  $\psi_0$
goes to the trivial vacuum as $x_0\to\infty$, and becomes a widely
separated soliton/antisoliton pair as $x_0\to 0$, crossing $\phi_0$ at
$x_0 \approx 0.75$.  For $x_0$ below this value, $\phi_0$ has lower
classical energy, while above this point, $\psi_0$ has lower classical
energy, and this crossing appears precisely where the imaginary
eigenvalue appears in the small oscillations spectrum.

Through this continuous deformation from the trivial vacuum, we have
arrived at a widely separated soliton/antisoliton pair, which we can
now separate into independent configurations with nontrivial
topology.  These configurations are exactly soluble, so we will be
able to study them analytically.  However, our method does not rely on
having an analytic solution, so we could numerically calculate the energy of a
generic field configuration with nontrivial topology using the same techniques.

In one dimensional quantum mechanics, potentials of the form
\begin{equation}
V_\ell(x) = -\frac{\ell+1}{\ell}m^2{\rm sech}^2(\frac{mx}{\ell})
\end{equation}
with $\ell$ an integer are exactly soluble and reflectionless.  The single
soliton solution in our model, $\phi_0(x) = \tanh\frac{mx}{2}$,
corresponds to $V(x) = -\frac{3}{2}m^2{\rm sech}^2\frac{mx}{2}$, the
$\ell=2$ case of this family. (The sine-Gordon soliton corresponds to the
$\ell=1$ case.) For a reflectionless potential, $\delta_{\rm S}(k) =
\delta_{\rm A}(k)$; to reconcile this equality with Levinson's theorem in
the symmetric and antisymmetric channels, $V_l(x)$ must have a half-bound
state.  (We saw this behavior already for the $\ell = 0$ case, the free
particle.)  We also note that although  $\delta_{\rm S} = \delta_{\rm A}$, 
$\delta^{(1)}_{\rm S} \neq \delta^{(1)}_{\rm A}$, so the renormalized
contributions from the symmetric and antisymmetric channels are not the
same.  In addition, the bound state contributions will also be unequal.

For our  case, the exact result for the phase shift is
\begin{equation}
\delta_{\rm S}(k) = \delta_{\rm A}(k) =  -\arctan \frac{3mk}{m^2-2k^2},
\end{equation}
where the branch of the arctangent is chosen so that the phase shift is
continuous and goes to zero for $k\to\infty$.  The Born approximation is
\begin{equation}
\delta^{(1)}_{\rm S}(k) + \delta^{(1)}_{\rm A}(k) = \frac{3m}{k}.
\end{equation}

There are three bound states: a translation mode with $E=0$, a state with
$E = \frac{m\sqrt{3}}{2}$, and a half-bound state with $E=m$.
Using eq.~(\ref{renorm1}) or eq.~(\ref{renorm2}) we find
\begin{equation}
{\cal E}[\phi_0] = m(\frac{1}{4\sqrt{3}} - \frac{3}{2\pi})
\end{equation}
in agreement with \cite{DHN}, \cite{Bordag}, and the
mode number cutoff method in \cite{vN}.

We can calculate the energy of the soliton of the sine-Gordon model using
the same methods.  In this model, the $\phi^4$ potential is replaced by
$U(\phi) = m^2(\cos\phi - 1)$, which has soliton (anti-soliton) solutions
$\phi_0 = 4 \arctan(\exp(\pm m(x-x_0)))$.  The phase shift is 
\begin{equation}
\delta_{\rm S}(k) = \delta_{\rm A}(k) =  \arctan \frac{m}{k}
\end{equation}
with Born approximation
\begin{equation}
\delta^{(1)}_{\rm S}(k) + \delta^{(1)}_{\rm A}(k) =  \frac{2m}{k}.
\end{equation}
The bound states are just the translation mode at $E=0$ and the
half-bound state at $E=m$, giving
\begin{equation}
{\cal E}[\phi_0] = -\frac{m}{\pi}
\end{equation}
again agreeing with the established results.

\section{Conclusions}

We have demonstrated that our method allows us to calculate one-loop
energies of a wide variety of static, symmetric field configurations.  We
have avoided ever having to deal directly with subtleties of the cutoff or
the boundary conditions, relying instead on the formalism of the Born
approximation and Levinson's theorem.  Using this formalism, we can apply
renormalization conditions fixed in the topologically trivial sector so
that our theory gives unambiguous predictions in terms of well-defined
masses and coupling constants, and express the result in terms of simple
physical quantities that allow for robust numerical and analytic
calculation.


\section*{Acknowledgments}

We would like to thank E.~Farhi, J.~Goldstone, P.~Haagensen, and P.~van
Nieuwenhuizen for helpful conversations, suggestions and references.  We
are grateful to S.~Bashinsky for correcting our understanding of the
crossing of solutions as a function of $x_0$. We also thank M.~Stock for
technical help in preparing the manuscript. We thank L.-H.~Chan for finding
an error in our numerical calculations by comparing our results with those
of \cite{Chan}.  This work is supported in part by funds provided by the
U.S.  Department of Energy (D.O.E.) under cooperative research agreement
\#DF-FC02-94ER40818, and by the RIKEN BNL Research Center.  N.~G. is
supported in part by an NSF Fellowship.  R.~L.~J. is supported in part by
the RIKEN-BNL Research Center.




\begin{references}

\bibitem{us} E.~Farhi, N.~Graham, P.~Haagensen, and R.~L.~Jaffe,
hep-th/9802015.

\bibitem{vN}
A.~Rebhan and P.~van Nieuwenhuizen, Nucl.~Phys.~{\bf B508} (1997) 449.
H.~Nastase, M.~Stephanov, P.~van Nieuwenhuizen and A.~Rebhan, hep-th/9802074.

\bibitem{Naculich1d} S.~G.~Naculich, Phys.~Rev~{\bf D46} (1992) 5487.

\bibitem{Bordag} M.~Bordag, J.~Phys.~A: Math.~Gen.~{\bf 28} (1995) 755.

\bibitem{DHN}
R.~Dashen, B.~Hasslacher, and A.~Neveu, Phys.~Rev.~{\bf D10} (1974) 4114,
4130.

\bibitem{Cole}  S.~Coleman,
 {\sl Aspects of Symmetry} (Cambridge University Press, Cambridge, 1985);
(North-Holland, Amsterdam, 1982).

\bibitem{Schiff} L.~Schiff, {\sl Quantum Mechanics} (McGraw-Hill, NY, 1968).

\bibitem{lev1}  G. Barton, J.~Phys.~A: Math.~Gen.~{\bf 18} (1985) 479.

\bibitem{Chan}
L.-H.~Chan, Phys.~Rev.~{\bf D55} (1997) 6223.

\end{references}
\end{document}